\documentclass[12pt]{article}

\usepackage{amssymb}
\usepackage{amsmath}
\usepackage{amscd}
\usepackage{latexsym}
\usepackage{graphicx}

\usepackage{cite}

\newcommand{\be}{\begin{equation}}
\newcommand{\ee}{\end{equation}}

\newcommand{\dlt}{\delta}

\newcommand{\br}{{\bf r}}

\newcommand{\al}{\alpha}

\newcommand{\om}{\omega}

\begin{document}

\begin{center}

{\Large{\bf Characterization of nonequilibrium states of trapped Bose-Einstein 
condensates} \\ [5mm]

V.I. Yukalov$^{1,2,*}$\let\thefootnote\relax\footnote{$^*$Author to whom correspondence 
should be addressed (V.I. Yukalov) \\ E-mail: yukalov@theor.jinr.ru \\ 
Phone: +7 (496) 213$-$3824}, A.N. Novikov$^2$, and V.S. Bagnato$^2$  } \\ [3mm]

{\it

$^1$Bogolubov Laboratory of Theoretical Physics, \\
Joint Institute for Nuclear Research, Dubna 141980, Russia \\ [3mm]

$^2$Instituto de Fisica de S\~{a}o Calros, Universidade de S\~{a}o Paulo, CP 369,  \\
 S\~{a}o Carlos 13560-970, S\~{a}o Paulo, Brazil   

}

\end{center}

\vskip 3cm

\begin{abstract}

Generation of different nonequilibrium states in trapped Bose-Einstein condensates
is studied by numerically solving nonlinear Schr\"{o}dinger equation. Inducing 
nonequilibrium states by shaking the trap, the following states are created: weak 
nonequilibrium, the state of vortex germs, the state of vortex rings, the state of straight 
vortex lines, the state of deformed vortices, vortex turbulence, grain turbulence, and 
wave turbulence. A characterization of nonequilibrium states is advanced by 
introducing {\it effective temperature, Fresnel number, and Mach number}. 

\end{abstract}

\vskip 1cm


\vskip 1cm


\newpage

\section{Introduction}

The most widely studied nonequilibrium state of quantum fluids is quantum turbulence 
\cite{Vinen_1,Vinen_2,Vinen_3,Tsubota_4,Nemirovskii_5}. Recently, quantum turbulence 
has got a lot of attention in the studies of cold atomic systems, as can be inferred form  
the review articles \cite{Tsubota_4,Nemirovskii_5,Wilson_6,Tsatsos_7}. The creation 
of strongly nonequilibrium atomic states of Bose-Einstein condensates by shaking a trap, 
as suggested in Refs. \cite{Yukalov_8,Yukalov_37}, has made it possible to experimentally 
realize quantum turbulence of cold atomic gases in harmonic traps (see reviews 
\cite{Nemirovskii_5,Wilson_6,Tsatsos_7}) and in a box-shaped trap \cite{Navon_9}. 

But quantum turbulence is not the sole nonequilibrium state that can be generated by
shaking a trap. Other nonequilibrium states, such as grain turbulence and wave 
turbulence can be generated in experiment and analyzed numerically by solving the
nonlinear Schr\"{o}dinger (NLS) equation 
\cite{Bagnato_10,Yukalov_11,Yukalov_12,Yukalov_13}.  

An important question is: How would it be possible to ascribe a quantitative characteristic 
to different nonequilibrium states that could be generated in trapped Bose-condensed 
systems? For classical liquids, laminar and turbulent flows are distinguished by the 
Reynolds number $Re = v L/ \nu$, where $v$ is a characteristic flow velocity, $L$ is a 
characteristic linear system size, and $\nu$ is kinematic viscosity \cite{Salmon_14}. 
The characterization of flows by Reynolds numbers is also applicable to quantum liquids, 
such as superfluid helium \cite{Donnelly_15}. However, for cold atomic gases, at 
low temperatures, viscosity is zero, so that the Reynolds number becomes infinite 
\cite{Barenghi_16}. For describing the motion of obstacles through atomic superfluid gas, 
the Strouhal number $St = f L/ v$ can be used \cite{Reeves_17,Kwon_18}, where $f$ is 
the vortex shedding frequency. Various patterns of vortex shedding can arise in quantum 
fluids behind moving obstacles \cite{Sasaki_19,Stagg_20}.  

However, the problem of characterizing various nonequilibrium states that could arise in 
trapped atomic Bose-condensed systems, without any obstacles, remains unsolved. 

The aim of the present paper is twofold. First, we thoroughly analyze what qualitatively 
different nonequilibrium states can appear in trapped Bose-Einstein condensates when 
the trapping potential is perturbed by time-dependent modulation. Second, we suggest a
quantitative characterization of all such states by three quantities, {\it effective 
temperature}, {\it Fresnel number}, and {\it Mach number}.  

The use of the Fresnel number is motivated by the analogy between nonequilibrium states 
of atomic systems and optical phenomena in lasers, where the terms {\it optical turbulence} 
or {\it photon turbulence} are well known. The Mach number has also been used for 
characterizing turbulent flows in geophysics, as is discussed in the book by Smits and 
Dussauge \cite{Smits_36}. Fresnel numbers for trapped cold atoms have also been 
discussed in the context of superradiant Rayleigh scattering \cite{Inouye_43}. These 
analogies suggest that the mentioned characteristics can be useful for describing 
nonequilibrium states of trapped atoms.  
 
Our goal is to model the behavior of Bose-Einstein condensate by numerically solving the
three-dimensional time-dependent nonlinear Schr\"{o}dinger equation describing the 
system of Bose-condensed atoms \cite{Pethick_21}. The system parameters are chosen 
exactly coinciding with the recent experiments \cite{Shiozaki_22,Seman_23} with $^{87}$Rb. 
This choice allows us to compare our numerical results with the available experimental data. 
But such a comparison is not the aim of the present paper, since it has already been done 
in our previous publications \cite{Yukalov_11,Yukalov_12,Yukalov_13} and good agreement 
between numerical and experimental data has been observed. In the present paper, we 
concentrate on a more detailed numerical  investigation of arising nonequilibrium states, 
finding additional types of the states, such as vortex germs, vortex rings, and strongly 
deformed vortices that have not been observed in the earlier publications.  
\cite{Yukalov_11,Yukalov_12,Yukalov_13}. The most important point of the present paper
is the suggested novel characterization of nonequilibrium states. We illustrate this 
characterization by nonequilibrium states generated in trapped Bose-Einstein condensates.
But this characterization, being formulated in general terms, can be applied to any other 
nonequilibrium states.      

In the brief letter \cite{Yukalov_12} we studied experiments with nonequilibrium trapped atoms 
and the interpretation of the experimentally observed states. The discussed experiments 
exhibit only three such states: separate vortices, vortex turbulence, and grain turbulence. 
While in the present paper, we accomplish a detailed numerical investigation of all possible 
nonequilibrium states, distinguishing eight different states: weak nonequilibrium, vortex 
germs, vortex rings, vortex lines, deformed vortices, vortex turbulence, grain turbulence, 
and wave turbulence.  Also, the injected energy, considered in \cite{Yukalov_12}, turned 
out to be not the most convenient quantity, because of which we suggest here new 
characteristics.

Defining such dimensionless quantities as a Fresnel and Mach number is useful in that they 
allow for comparison between systems with different geometries and length scales and also 
between different physical systems. In this light, defining a Fresnel number for a BEC allows 
one to draw parallels between optical and atomic nonequilibrium states, e.g., such as 
turbulence.

The transitions between the observed nonequilibrium states are not sharp phase transitions,
but they are gradual crossovers. However, each state is qualitatively different from others,
because of which they can be well distinguished from each other and classified as separate
states. The gradual crossover transitions are typical for finite systems and for nonequilibrium 
phenomena \cite{Binder_46}.

\section{Generation of nonequilibrium states}

We  choose the same setup as has been used in the recent experiments 
\cite{Shiozaki_22,Seman_23} with $^{87}$Rb. At the initial moment of time, all atoms of  
$^{87}$Rb, with the mass $m = 1.445 \times 10^{-22}$g and the scattering length 
$a_s =  0.577 \times 10^{-6}$ cm, are assumed to be Bose-condensed in a cylindrical 
harmonic trap with the radial frequency $\omega_r = 2 \pi \times 210$ Hz and the axial 
frequency $\omega_z = 2 \pi \times 23$ Hz. The total number of atoms $N = 1.5 \times 10^5$. 
The atomic cloud has the radius $R_{eff} = 4 \times 10^{-4}$ cm and length 
$L_{eff} = 6 \times 10^{-3}$ cm. The central density is $\rho = 2.821 \times 10^{14}$ cm$^{-3}$. 
The healing length is $ \xi = 1.106 \times 10^{-5}$ cm. 

As usual, for numerical calculations, it is convenient to write the time-dependent equation 
in a dimensionless form, with the energies measured in units of the characteristic oscillator
frequency $\omega_0 \equiv (\omega_r^2 \omega_z)^{1/3}$, time measured in units 
of $\omega_0^{-1}$, and lengths measured in units of the characteristic oscillator length 
$l_0 \equiv \sqrt{\hbar / m \omega_0}$. Here $\omega_0 = 630$ s$^{-1}$ and 
$l_0 = 1.08 \times 10^{-4}$ cm.   

Starting from the initial time, the trap is subject to the action of a perturbing alternating 
potential, such that the trapping potential, in dimensionless units, takes the form
$$
V(\br,t) = \frac{1}{2} \left \{ \frac{1}{\al^2} \; [ x\cos\vartheta_1 + y \sin\vartheta_1  
- z\sin \vartheta_2 - \dlt_1 ( 1 - \cos \om t ) ] ^2 \right. +
$$
\be
\label{1}
 + \left.
[ y\cos\vartheta_1 - x \sin\vartheta_1   - \dlt_2 ( 1 - \cos \om t ) ] ^2 +
[ z\cos\vartheta_2 + x \sin\vartheta_2  - \dlt_3 ( 1 - \cos \om t ) ] ^2 \right \} \;.
\ee
Here $\alpha \equiv \omega_z/\omega_r$ is aspect ratio, 
$\vartheta_i = A_i(1 - \cos\om t)$, $A_1 = \pi/60$, $A_2 = \pi/120$, $\dlt_1 = 1.861 A$, 
$\dlt_2 = 4.653 A$, $\dlt_3 = 2.792 A$, the effective modulation amplitude $A = 0.2$, and 
the modulation frequency is $\omega = 2 \pi \times 200$ Hz. These parameters could be 
varied, which, as we have checked, does not qualitatively change the overall picture.
So, in numerical calculations, we keep them as defined above. The alternating trapping 
potential shakes the atomic cloud without imposing a moment of rotation. 

Dynamics of the trapped Bose condensate is analyzed by numerically solving the nonlinear 
Schr\"{o}dinger (NLS) equation defined on a three-dimensional Cartesian grid. The grid 
contains $2^{30}$ points, its mesh size is approximately the half of the healing length. This 
set of parameters allows us to avoid the unphysical reflection from the grid boundaries
and properly simulate all nonequilibrium states. Energy dissipation is taken into account by 
introducing a phenomenological imaginary term into the left-hand side of the NLS equation.
For more detailed information on the numerical procedure, we refer to Ref. \cite{Novikov_24}. 

In the process of perturbation, there appear different spatial structures, such as vortex 
germs, vortex rings, vortex lines, and grains. The lifetime of these structures is defined 
in the following way.  After they are created, the external pumping is switched off and 
the behavior of the structures in the stationary trap is monitored. 

The sequence of the observed nonequilibrium states is as follows.

(i) {\it Weak nonequilibrium}. At the first stage of the process, lasting around $5$ ms, 
the energy injected into the system is not yet sufficient for generating topological modes,
but  produces only density fluctuations above the equilibrium state of the atomic cloud. 
 
(ii) {\it Vortex germs}. After  $5$ ms of perturbation, at the low-density edges of the 
cloud, there appear pairs of vortex germs reminding broken pieces of vortex rings. 
These objects do possess vorticity $\pm 1$, which is found by reconstructing vortex 
lines in the coordinate space employing the method described in Refs. 
\cite{Tsubota_4,Kobayashi_35}. According to this method, the points are defined 
representing the location of phase defects in the three-dimensional coordinate space 
and thus defining the position of the vortex core. It is assumed that the vortex line 
passes through the nearest points of the phase defects. If the external pumping is 
switched off  after the germs are created, they survive during about $0.2$ s, when 
they do not move, except exhibiting small oscillations. But if we continue pumping 
the energy into the trap, we proceed to the next stage. Typical vortex germs are shown 
in Fig. 1, where the related modulation time is marked. 

\begin{figure}
\begin{center}
\includegraphics[width=120mm,height=75mm]{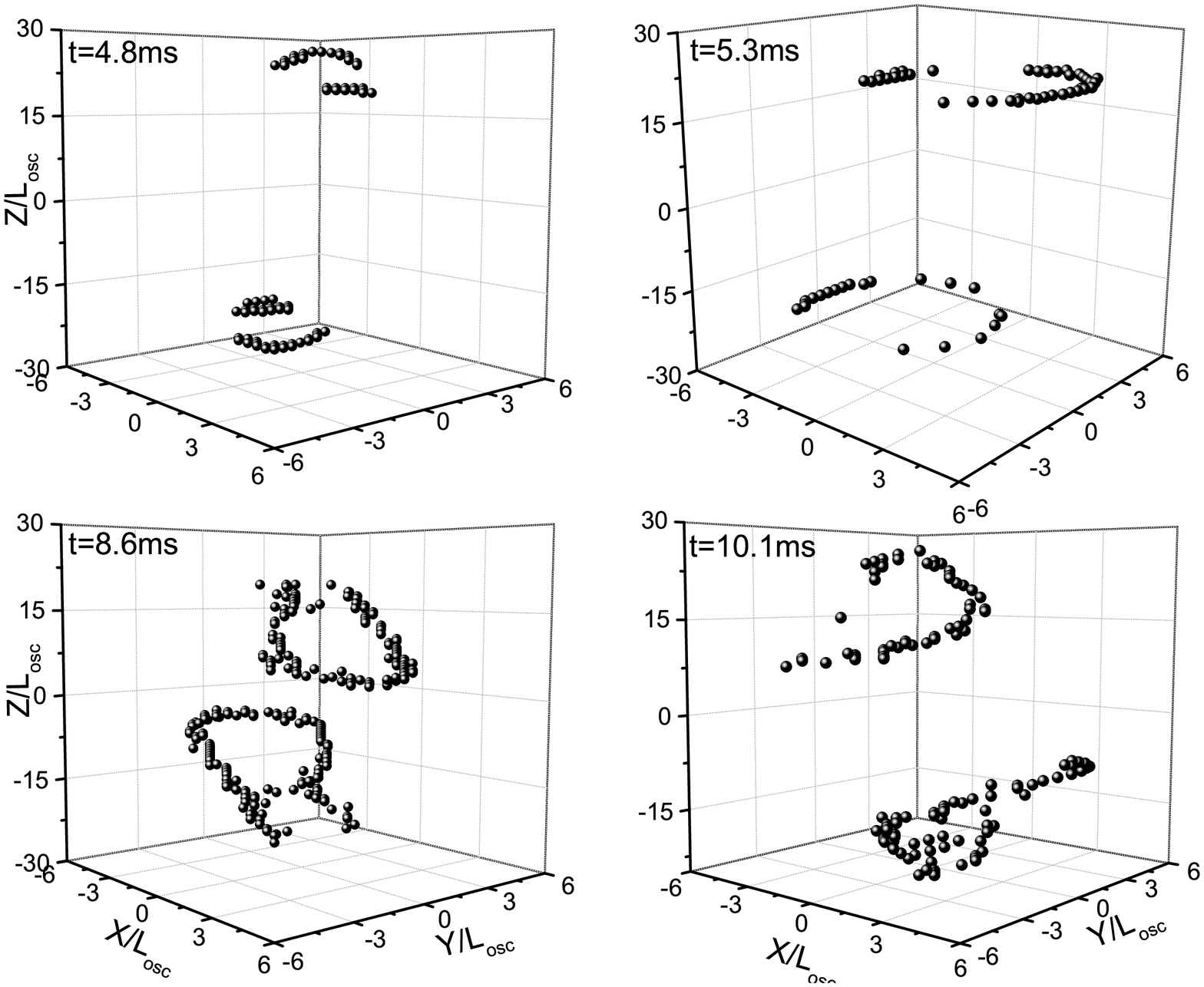}
\end{center}
\caption{Typical examples of vortex germs. Here and in the following figures
the time of the trap modulation is marked.}
\end{figure}

(iii) {\it Vortex rings}. After around $10$ ms, instead of the vortex germs, there appear
well defined pairs of vortex rings, with vorticity $\pm 1$. Typical examples of vortex rings 
are illustrated by Fig. 2, where the modulation time is also marked. The rings do not move, 
except exhibiting small oscillations. The ring lifetime, after switching off pumping, is about 
$0.1$ s.  

(iv) {\it Vortex lines}. Continuing pumping energy into the system leads, after about 
$15$ ms, to the formation of pairs of straight vortex lines, with vorticity $\pm 1$, 
directed along the $z$-axis, as is demonstrated in Fig. 3. The lines randomly move, 
probably, due to the Magnus force. The vortex lifetime, after switching off pumping, is 
about $0.2$ s.

(v) {\it Deformed vortices}. The longer perturbation, after around $17$ ms, starts
strongly deforming vortex lines, making them not straight and chaotically directed.
The lifetime of the deformed vortices is close to that of the straight vortex lines.

\begin{figure}
\begin{center}
\includegraphics[width=120mm,height=75mm]{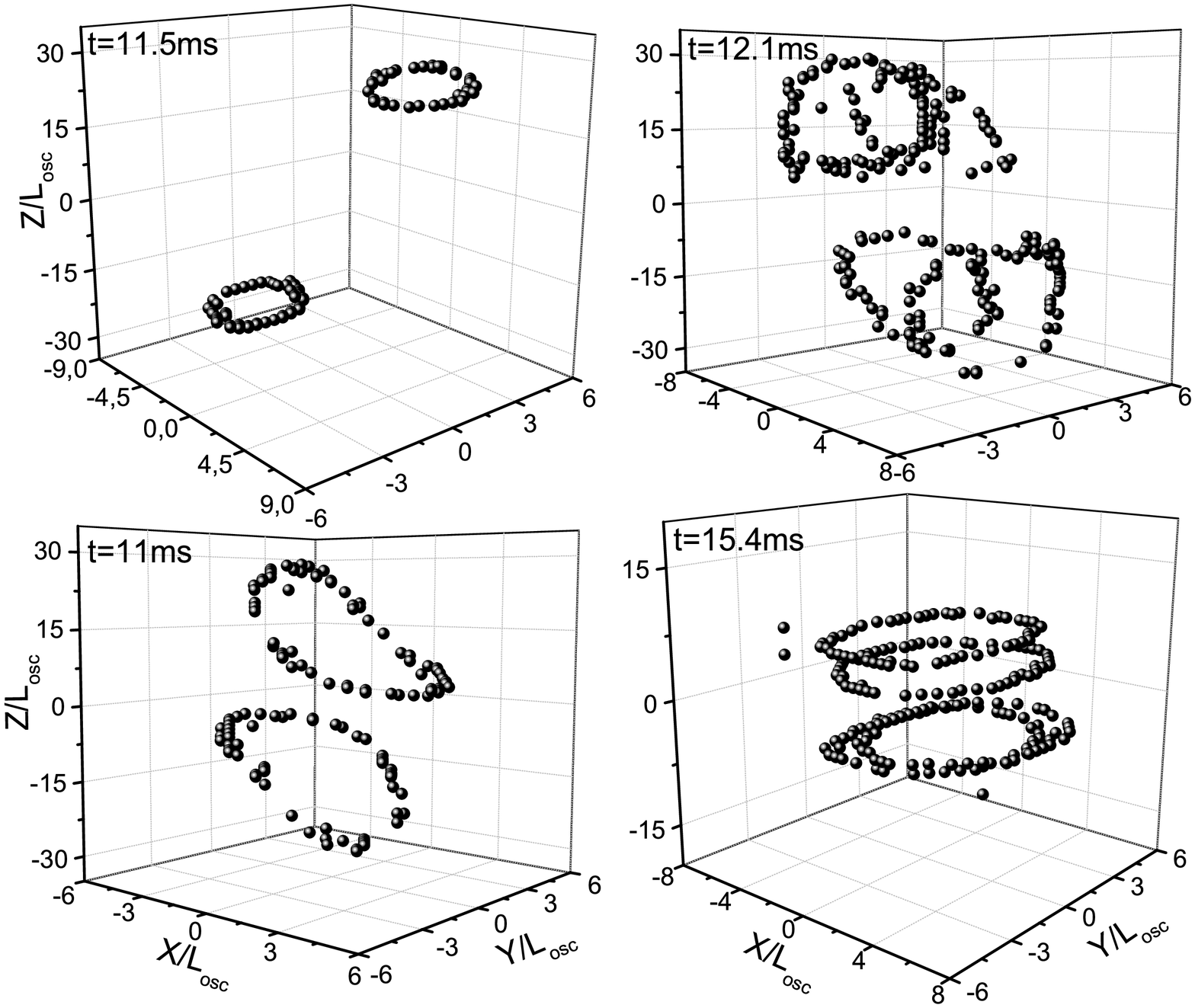}
\end{center}
\caption{Typical examples of vortex rings.}
\end{figure}

\begin{figure}[b]
\begin{center}
\includegraphics[width=120mm,height=50mm]{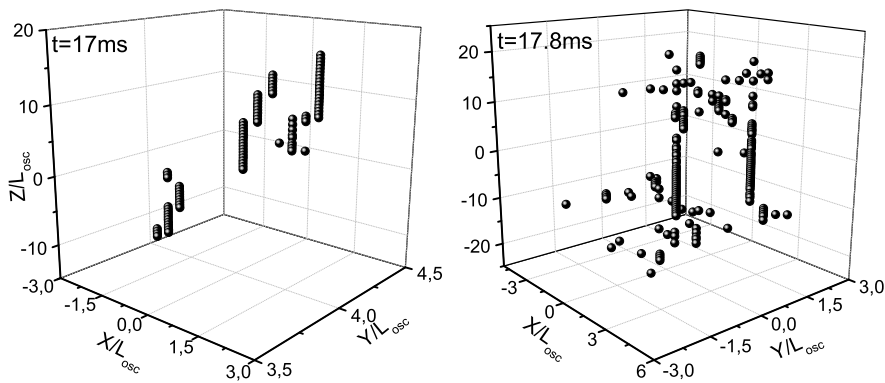}
\end{center}
\caption{Typical examples of straight vortex lines directed along the $z$ axis.}
\end{figure}

(vi) {\it Vortex turbulence}. After about $25$ ms, the number of deformed vortices 
sharply increases and they form a randomly oriented vortex tangle, typical of the 
Vinen turbulence \cite{Vinen_2,Vinen_3,Tsubota_4,Cidrim_44}. The total regime 
of vortex turbulence can be subdivided onto the stages of developing turbulence,
developed turbulence, and decaying turbulence. The stage of developed vortex 
turbulence is demonstrated in Fig. 4. The regime of vortex turbulence can be 
classified as such, since it exhibits several features typical of quantum turbulence:
First, there appears a random tangle of vortices, corresponding to the 
definition of quantum turbulence by Feynman \cite{Feynman_25}. Second, being
released from the trap, the atomic cloud expands isotropically, which is typical of
Vinen turbulence \cite{Caracanhas_26}. Third, the column-integrated radial 
momentum distribution obeys a power low $n_r(k) \propto k^{-\gamma}$ in the 
range \cite{Tsubota_4} $k_{min} < k < k_{max}$, with $k_{min} \approx \pi / \xi$ 
and $k_{max} \approx 2 \pi / \xi$, which is a key quantitative expectation for an 
isotropic turbulent cascade \cite{Zakharov_27}, and which has been observed 
for turbulence of trapped atomic clouds \cite{Navon_9,Thompson_28}. We have 
found that in this range $0.284 \times 10^6 \rm{cm}^{-1} < k < 0.568 \times 10^6 \rm{cm}^{-1}$,
the slope of $n_r(k)$ is $\gamma \approx 1.7$, which is close to $\gamma \approx 2$ 
found in the experiment \cite{Thompson_28} accomplished for the same system 
with the same parameters.

\begin{figure}[t]
\begin{center}
\includegraphics[width=120mm,height=50mm]{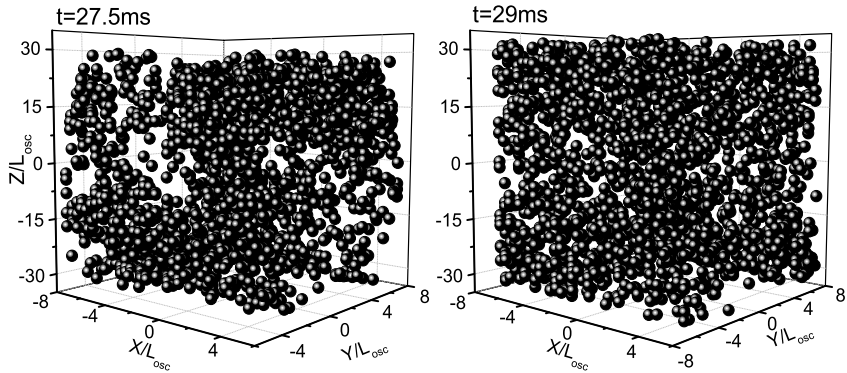}
\end{center}
\caption{Developed vortex turbulence.}
\end{figure}

(vii) {\it Grain turbulence}. The regime of decaying vortex turbulence, after 
approximately $40$ ms, is followed by the state, where there are practically no 
vortices, but the whole trap is filled by randomly distributed in space dense 
Bose-condensed droplets, or grains, surrounded by a rarified gas. The droplets are 
almost spherical, with the radii $(0.5 - 2.5) \times 10^{-5}$ cm. The typical droplet 
radius of $1.5 \times 10^{-5}$ cm is close to the coherence length, as it should be. 
The density inside a droplet is $20 - 100$ times higher than in the surrounding gas. 
The lifetime of a grain, after switching off pumping, is about $0.01$ s. During this time, 
droplets chaotically move in space, then some of them disappear, while new appear.
The droplet lifetime is much longer that the local equilibration time 
$t_{loc} = m / \hbar \rho a_s = 0.8 \times10^{-3}$ s. Hence the droplets are metastable 
objects. Each droplet is coherent, having a constant phase, while in the surrounding 
the phase is random.  Such a state can be treated as a heterophase mixture of 
Bose-condensed droplets immersed into the gas of uncondensed atoms 
\cite{Yukalov_29,Yukalov_45}. The density snapshot for a radial cross-section, comparing 
the state of grain turbulence with the equilibrium unperturbed condensate and with the 
following state of wave turbulence is shown in Fig 5. It is important to stress that 
the granulated state of matter, to be classified as such, has to satisfy the following
criteria: (1) the typical size of each droplet, representing a coherent formation, is of 
the order of the healing length; (2) the phase inside a droplet is constant, as it should 
be for a coherent object; (3) the phase in the space around a droplet is random, so that
the coherent droplets are separated from each other by incoherent surrounding; 
(4) the lifetime of a droplet is much longer than the local equilibration time, which defines
the droplets as metastable objects; (5) to look really as a droplet inside a rarified gas, 
the density inside a droplet is to be much larger than that of its incoherent surrounding.
If these criteria are not satisfied, the matter cannot be classified as granulated. In our 
case all these criteria hold true. 
    
(viii) {\it Wave turbulence}. Increasing perturbation by the trap modulation destroys
Bose-condensed droplets after about $150$ ms and the system transfers into the
state of wave turbulence \cite{Nazarenko_30,Fujimoto_31}. The state is a collection
of small-amplitude waves, with the sizes $(0.5 - 1.5) \times 10^{-4}$ cm and with the 
density only about $3$ times larger than the density of their surrounding. There is 
no coherence either inside a wave or between them, since the phase is random. 
Strictly speaking, the transition from the regime of grain turbulence to wave turbulence 
is a gradual crossover, with more and more destroyed coherence. The latter 
becomes practically completely destroyed, so that the phase is everywhere chaotic, 
at about  $150$ ms. The density snapshot of the state is demonstrated in Fig. 5.

\begin{figure}
\begin{center}
\includegraphics[width=160mm,height=50mm]{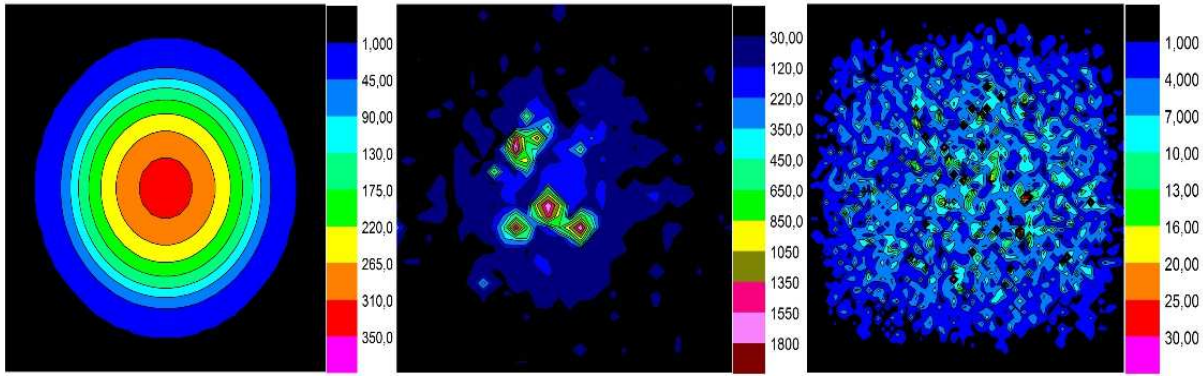}
\end{center}
\caption{Density snapshots of equilibrium condensate (left plot), grain turbulence
(middle plot), and wave turbulence (right plot). The legends on the right-hand side
of each plot show the density in units of $1 / l_0^3$.}
\end{figure}

\section{Quantitative characterization of nonequilibrium states}

In this way, by injecting energy into the trap, we create different nonequilibrium states
with various topological defects, finally transforming an initially equilibrium Bose-condensed 
system into the state, where Bose-Einstein condensate is completely destroyed. This 
process, being opposite to equilibration, can be treated as inverse Kibble-Zurek scenario 
\cite{Yukalov_13}. The remaining problem is how it would be possible to quantitatively 
characterize the nonequilibrium states arising during this disequilibration process. 
Several characteristics can be suggested for such a characterization. 

First of all, it is possible to check that the energy, pumped into the system, goes almost
completely to the increase of kinetic energy \cite{Yukalov_13}. The latter can be connected 
with the effective temperature. Therefore, it is natural to define the {\it effective temperature},
in energy units, by the increase of kinetic energy per atom from the initial value $E_{kin}(0)$ 
to its value $E_{kin}(t)$ at time $t$,
\be
\label{2}
T_{eff}(t) \equiv \frac{2}{3} \; [ E_{kin}(t) - E_{kin}(0) ]  \;   .   
\ee
Note that this expression is written in energy units. To get the units of Kelvin temperatures,
one has to divide Eq. (\ref{2}) by the Boltzmann constant $k_B$. 

The other idea comes to our mind if we remember that the phenomenon of turbulence
is very general, arising in different media. Stochastic vortex tangles exist in many 
physical fields of similar systems of highly disordered sets of one-dimensional topological 
objects. As examples, it is possible to mention global cosmic strings, the flux tubes in 
superconductors, dislocations in solids, linear topological defects in liquid crystals and 
polymer chains, turbulent effects in quark-gluon plasma and neutron stars (see discussion 
in reviews \cite{Tsubota_4,Nemirovskii_5}). The emergence of multiple random filamentation 
in a high-intensity, ultrashort laser beam, due to optical turbulence, has been observed 
\cite{Berge_38,Henin_39,Loriot_40,Ettoumi_41}. The phenomenon of {\it photon turbulence}, 
or {\it optical turbulence}, also occurs in high-Fresnel-number lasers, where the increase 
of the laser Fresnel number leads to the appearance of turbulent photon filamentation 
\cite{Yukalov_32,Yukalov_33,Yukalov_34}. 

That is, the Fresnel number can serve as a characteristic of the system state. For lasers, 
with wavelength $\lambda$, radius $R$, and length $L$, the Fresnel number is 
$F = \pi R^2/ \lambda L$. For atomic traps, the role of radius is played by the oscillator 
length $R = l_r = \sqrt{\hbar/ m\omega_r}$, while the axial length is given by 
$L = 2 l_z = 2 \sqrt{\hbar / m \omega_z}$. As the wavelength, it is natural to accept the 
length $\lambda_T = \hbar \sqrt{ 2 \pi / m T_{eff}}$. Thus, the effective Fresnel number 
can be written as
\be
\label{3}
 F =  \frac{\pi R^2}{\lambda_T L}  =  \sqrt{ \frac{\pi \al T_{eff}}{8\hbar\om_r} } \;  ,
\ee
where $\alpha$ is the aspect ratio.

It is also possible to recollect that turbulence is connected with such a characteristic as
the Mach number $M = v / c$, in which $v$ is the velocity of a moving object and $c$
is sound velocity \cite{Smits_36}. For atomic systems, the velocity of an atom can be 
represented as $v = \sqrt{2 E_{kin} / m}$ and the sound velocity can be written as 
$c = (\hbar / m) \sqrt{4 \pi \rho a_s}$, where $\rho$ is the density at the trap center.
Hence the effective Mach number takes the form
\be
\label{4}
 M = \sqrt{\frac{2E_{kin}}{mc^2}} =  \frac{1}{\hbar} \; \sqrt{ \frac{m E_{kin}}{2\pi\rho a_s} } \;   .
\ee
The first expression here is what is called {\it turbulence Mach number} \cite{Smits_36}.

In general, the Mach number shows the relation between the object velocity and the 
speed of sound. Considering an obstacle, such as, e.g., laser beam moving through 
the fluid, is a particular case. In our study the Mach number characterizes the relation 
between the characteristic velocity of atoms and the speed of sound. The Mach number 
we introduce describes how a particle moves inside the system. In that sense, a particle 
is also an effective moving object, similar to an obstacle. So, the suggested definition 
is a straightforward generalization of the standard Mach number. Recall that cold atoms, 
we consider, are not thermal. Their speed has no connection with thermal motion, but 
reflects the speed due to quantum kinetic energy. In the case of cold atoms, kinetic energy 
is caused by quantum motion, while for classical systems kinetic energy is really thermal. 
This makes the principal difference between quantum and classical systems.  

In Fig. 6, we illustrate the classification of all nonequilibrium states, we have found,
by means of the effective temperature, Fresnel number, and Mach number. It is 
interesting that the regime of wave turbulence, where all coherence is destroyed, 
corresponds to the effective temperature $T_{eff} = 23.5 \hbar \omega_r$, which 
practically coincides with the temperature $T_c = 23.8 \hbar \omega_r$ of 
Bose-Einstein condensation of $^{87}$Rb in the considered setup. The critical 
temperature in Kelvin degrees is $2.4 \times 10^{-7}$ K.  The developed wave 
turbulence implies that, although the system is yet quantum, but there is no coherence, 
as it should be above the condensation temperature. The Fresnel and Mach numbers 
vary between small values close to zero, in weak nonequilibrium, to the values close 
to one, in wave turbulence.   

Note that for a strongly nonequilibrium system, the standard notion of temperature is 
not defined, but one can speak only about an {\it effective temperature}, as we do. The
closeness of the effective temperature at the moment, when the coherence is destroyed
by perturbations, to the Bose condensation temperature implies that coherence can be 
destroyed either by heating of an equilibrium system or by injecting kinetic energy into 
a nonequilibrium system, for which heating has no meaning.   

\begin{figure}[t]
\begin{center}
\includegraphics[width=140mm,height=80mm]{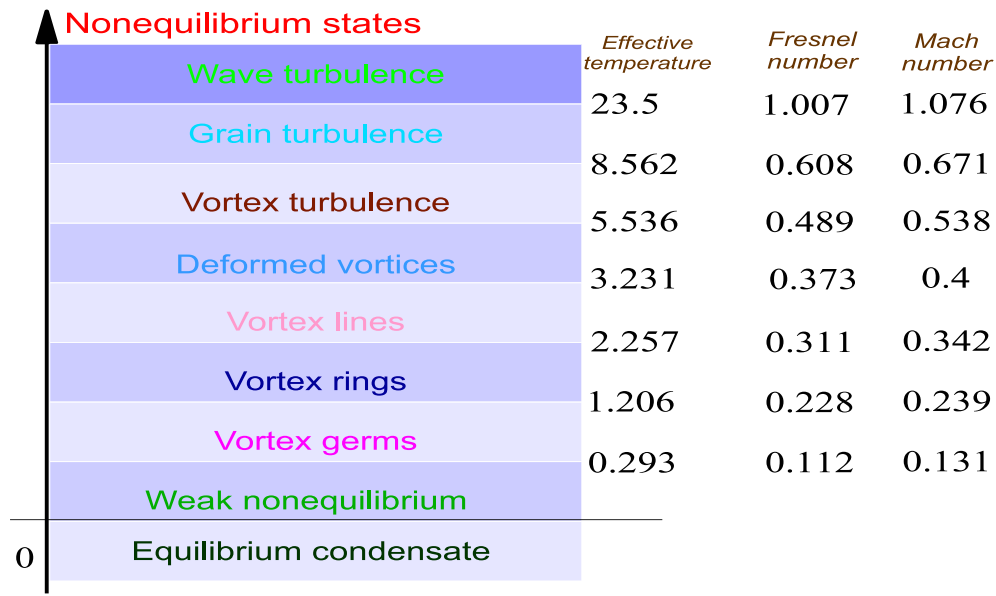}
\end{center}
\caption{Sequence of nonequilibrium states observed in computer modeling,
characterized by the effective temperature $T_{eff}$ (in units of $\hbar \omega_r$), 
Fresnel number, and Mach number.}
\end{figure}

\section{Conclusion}

In conclusion, we have accomplished a detailed numerical investigation of 
nonequilibrium states arising in the process of perturbing the Bose-Einstein condensate 
of trapped atoms by an alternating potential. Starting from an equilibrium condensate, 
we generate the state of weak nonequilibrium, the states with vortex germs, vortex
rings, vortex lines, and with deformed vortices, vortex turbulence, grain turbulence, and 
wave turbulence. The characterization of the nonequilibrium states is suggested by means
of effective temperature, Fresnel number, and Mach number. The latter are well defined
quantities expressed through the known system parameters and kinetic energy that can 
be straightforwardly calculated numerically as well as measured experimentally. The 
overall physical picture remains the same, if the perturbation parameters are varied. 
This is because the main characteristics of the states depend on the injected kinetic 
energy that can be shown \cite{Yukalov_11,Yukalov_12,Yukalov_13} is proportional to 
the product $A \omega t$ of the modulation amplitude $A$, modulation frequency 
$\omega$, and modulation time $t$. So, the same amount of kinetic energy can be 
injected into the trap by increasing the modulation amplitude, but decreasing the 
modulation time or modulation frequency. 

The suggested characteristics can be used for the traps of any geometry and for
any type of cold atoms. The main definitions of the effective temperature, Fresnel 
number, and Mach number remain the same, although the expressions for the  
system sizes $R$ and $L$, as well as for the sound velocity $c$, can be different, 
depending on the trap geometry and system parameters. For example, in the case 
of a cylindrical box \cite{Navon_9}, the system sizes are given by the radius $R$
and length $L$ of the box. 

It is important to emphasize that the characterization of nonequilibrium states, proposed
in the present paper is very general and can be employed for all systems for which 
such straightforward quantities as kinetic energy and the system sizes can be defined. 
These quantities, as is evident, can be defined for practically any experimental scheme.  
The suggested characteristics can be measured in experiments, provided kinetic energy 
can be measured. The latter can be found by measuring the momentum distribution $n({\bf k})$, 
as has been done in the experiments with trapped atoms \cite{Thompson_28,Bahrami_42}, 
after which the kinetic energy is straightforwardly obtained by integrating over momenta
of the expression $(k^2 / 2m) n({\bf k})$.  

\vskip 2mm

The authors are grateful to M. Tsubota for discussions and advice. One of the authors 
(A.N. N.) would like to thank CNPq (project 150343/2016-7) for post-doctoral fellowship
and  BLTP JINR for generous access to computational facilities. The other author (V.I.Y.)
appreciates discussions with E.P. Yukalova. The authors acknowledge useful cooperation 
with Center CEPOF and support from FAPESP (CEPID).

\end{document}